\documentclass[twocolumn]{autart}    % Enable this line and disable the
\usepackage{graphicx}
\usepackage{mathrsfs}
\usepackage{amsmath}
\usepackage{amssymb}          % Include this line if your
\begin{document}

\begin{frontmatter}

\title{Event-triggered consensus of multi-agent systems under directed topology based on periodic sampled-data\thanksref{footnoteinfo}} % Title, preferably not more

\thanks[footnoteinfo]{This paper was not presented at any IFAC
meeting. Corresponding author K.~E.~Liu.}

\author[a]{Kaien Liu}\ead{kaienliu@pku.edu.cn},    % Add the
\author[b]{Zhijian Ji},               % e-mail address
\author[c]{Xianfu Zhang}
%\author[Baiae]{Publius Maro Vergilius}\ead{vergilius@culture.ir}  % (ead) as shown

\address[a]{School of Mathematics and Statistics, Qingdao University, Qingdao, Shandong 266071, China}  % Please supply
\address[b]{College of Automation and Electrical Engineering, Qingdao University, Qingdao, Shandong 266071, China}             % full addresses
\address[c]{School of Control Science and Engineering, Shandong University, Jinan, Shandong 250061, China}
%Institute of Complexity Science, College of Automation and Electrical Engineering, Qingdao University, Qingdao, China
%\address[Baiae]{The White House, Baiae}        % here.

                             % keyword list or with the
                                          % help of the Automatica
                                          % keyword wizard

\begin{abstract}                          % Abstract of not more than 200 words.
The event-triggered consensus problem of first-order multi-agent systems under directed topology is investigated. The event judgements are only implemented at periodic time instants. Under the designed consensus algorithm, the sampling period is permitted to be arbitrarily large. Another advantage of the designed consensus algorithm is that, for systems with time delay, consensus can be achieved for any finite delay only if it is bounded by the sampling period. The case of strongly connected topology is first investigated. Then, the result is extended to the most general topology which only needs to contain a spanning tree. A novel method based on positive series is introduced to analyze the convergence of the closed-loop systems. A numerical example is provided to illustrate the effectiveness of the obtained theoretical results.
\end{abstract}

\end{frontmatter}

%======================================================================
\section{Introduction}
%======================================================================
As one of the fundamental problems of cooperative control in networks of multi-agent systems, the consensus problem has been extensively investigated in the past decade for its wide applications \cite{2012CaoY1,2014GaoY1,2014JiZ1,2015KLiu1,2018LiuX1}. In the existing literature, most of the designed consensus algorithms rely on continuous availability of communication information and the continuous control updates are applied. However, such manners may waste unnecessary energy and lead to delays because of network congestion for communication bandwidth \cite{2017KaienLiu,2012KELiu1,2014KELiu1,2018LiuX}.

Recently, the event-triggered control mechanism has been introduced in order to overcome the mentioned drawbacks. That is, consensus algorithms are designed according to the information at certain time instants determined by certain events and each agent will not change its control input until the new happenings of itself and its neighbors' events. In \cite{2012DimarogonasD1}, Dimarogonas et al. proposed several event-based consensus algorithms for first-order multi-agent systems, whose updates depended on the ratio of a certain measurement error with respect to the norm of a function of the states. In \cite{2013FanY1}, a combinational measuring approach to event-design was proposed. In \cite{2013Geoge1}, Seyboth et al. proposed an event-based scheduling strategy which bounded each agent's measurement error by a time-dependent threshold to solve average consensus. \cite{2016LiuK1} extended the method in \cite{2013Geoge1} to investigate the containment control problem of multi-agent systems under directed topology. Decentralised event-triggered consensus of double integrator multi-agent systems with packet losses and communication delays was investigated in \cite{2016Garcia1}. The event-based consensus problem for general linear multi-agent systems under directed topology was investigated in \cite{2014ZhuW}. Event-triggered leader-follower tracking control for multi-agent systems with general linear dynamics was considered in \cite{2016ChengY1}. \cite{2016LiH1} investigated event-triggered consensus of multi-agent systems with nonlinear dynamics and directed network topology. In the above works, though the intermittent control updates were applied, the continuous availability of communication information was still required in the event judgements. In \cite{2013MengX1} the average consensus problem was considered for multi-agent networks over undirected and connected topologies. The event-triggered condition was based on sampled-data in the sense that the event detector used only information at discrete sampling instants. Periodic event-triggered average consensus over strongly connected and balanced directed topology was considered in \cite{2016NowzariC1}. Effective delay-robust event-triggered consensus strategies, in which periodic event detectors were used to monitor and verify event-triggered conditions, were given in \cite{2017KaienL} and \cite{2016WangA1}, respectively. Asynchronous periodic event-triggered consensus for multi-agent systems was discussed in \cite{2017MengX1} and \cite{2017WangA1}, respectively. In \cite{2015CaoMT1}, consensus based on sampled-data control and edge event-driven techniques was investigated for second-order systems.  Event-triggered sampled-data consensus for multi-agent systems with general linear dynamics was considered in \cite{2014GuoG1}. The periodic sample and event hybrid control method used in \cite{2015CaoMT1,2014GuoG1,2017KaienL,2013MengX1,2017MengX1,2016NowzariC1,2016WangA1,2017WangA1} has obvious advantages. First, it can automatically rule out Zeno behavior which is often a challenge in distributed event-triggered systems. Second, the continuous availability of communication information is not required in the event judgements any more. Note that most of the above mentioned results focused on connected undirected topologies or strongly connected directed topologies. Moreover, the delay-robust consensus problem of multi-agent systems was only considered in \cite{2016Garcia1}, \cite{2017KaienL} and \cite{2013Geoge1,2016WangA1}.

Motivated by the above mentioned work, we discuss the consensus problem of multi-agent systems without or with time delay under directed topology. Our contributions include the design and analysis of novel event-triggered consensus algorithm to solve the consensus problem.
\begin{itemize}
\item First, under the designed consensus algorithm, the sampling period is permitted to be arbitrarily large, which avoids the requirement for the sampling period as in the existing results. Moreover, a corresponding advantage is that, for systems with time delay, consensus can be achieved for any finite delay only if it is bounded by the sampling period.\\
\item Second, compared with the existing results in \cite{2017KaienL,2013MengX1,2016NowzariC1,2016WangA1}, the topology required to guarantee consensus of multi-agent systems is extended to the most general case, i.e., with a spanning tree.
\end{itemize}

To prove the corresponding conclusions, the case of a strongly connected topology is first investigated. Then, the result is extended to the case of a topology containing a spanning tree. A novel method based on positive series is introduced to analyze the convergence of the closed-loop systems.

\sl Notations: \rm The following notations are used throughout this paper.
$\mathcal{I}_n=\{1,\ldots,n\}$ indicates an index set. $N$ and $Z^+$ denote the sets of all the nonnegative integers and all the positive integers, respectively.  Denote the $n\times 1$ column vectors of all ones and all zeros, respectively, as $\textbf{1}_n$ and $\textbf{0}_n$. $\mathbf{0}$ indicates a zero matrix with proper order.
%======================================================================
\section{Preliminaries}
%======================================================================
In this section, we first give a brief review of graph theory. Then, the model is formulated.

%\subsection{Graph Theory}
%----------------------------------------------------------------
%----------------------------------------------------------------
A {\it directed graph (digraph)} $G=(\mathscr{V},\mathscr{E})$ consists of a {\it node set} $\mathscr{V}=\{v_1,\ldots,v_n\}$, an {\it edge set} $\mathscr{E}\subseteq \mathscr{V}\times \mathscr{V}$. An {\it adjacency matrix} $A=[a_{ij}]_{n\times n}$ of $G$ is defined as $a_{ij}=1$ if $(v_i,v_j)\in \mathscr{E}$, while $a_{ij}=0$, otherwise. Here, we assume that $(v_i,v_i)\notin \mathscr{E}$ and hence $a_{ii}=0$ for $i\in\mathcal{I}_n$. The {\it in-degree} of node $v_i$ is defined as follows: $d_i=\sum_{j=1}^na_{ij}$, $i\in\mathcal{I}_n$. The {\it degree matrix} of $G$ is a diagonal matrix $D=\mathrm{diag}\{d_1,\ldots,d_n\}$. The {\it Laplacian matrix $L=[l_{ij}]_{n\times n}$} of $G$ is defined as $L=D-A$. Obviously, $d_i=l_{ii},\ i\in\mathcal{I}_n$. If there exists an edge $(v_i,v_j)\in \mathscr{E}$, then we say that node $v_j$ is a neighbor of node $v_i$. A {\it directed path} from $v_j$ to $v_i$ in a digraph is a sequence of edges starting with $v_j$ and ending with $v_i$. A digraph is called {\it strongly connected} if there exists a directed path from every node to every other node. A digraph contains a {\it spanning tree} if there exists a node called {\it the root node} such that there exists a directed path from it to every other node. An {\it undirected graph} means $(v_i,v_j)\in \mathscr{E}$ iff $(v_j,v_i)\in \mathscr{E}$. An undirected graph is called {\it completely connected} if any two distinct nodes are linked by an edge.

The following lemma can be deduced by combining Lemma 1 in \cite{2006LuW1}, Theorem 7 in \cite{2004Olfati-SaberR1}, and Lemma 3.3 in \cite{2005RenW2}.
%----------------------------------------------------------------
\begin{lem}\label{lm.1}
\ \\
(i) If a digraph $G$ contains a spanning tree, then zero is an algebraically simple eigenvalue of $L$ and all the other eigenvalues have positive real parts. When $G$ is strongly connected, $L$ has a positive left eigenvector $\xi^T=(\xi_1,\ldots,\xi_n)$ with $\sum_{i=1}^n\xi_i=1$ associated with zero eigenvalue. Moreover, $R=\frac{1}{2}(\Xi L+L^T\Xi)$ is a positive semi-definite matrix with all row sums being zeros and zero being its algebraically simple eigenvalue, where $\Xi={\rm diag}\{\xi_1,\cdots,\xi_n\}$.\\
(ii) Denote $U=\Xi-\xi\xi^T$.  A direct calculation gets that $(x-\sum_{i=1}^n\xi_ix_i\mathbf{1}_n)^T\Xi(x-\sum_{i=1}^n\xi_ix_i\mathbf{1}_n)=x^TUx$, where $x\in R^n$. Furthermore, $U$ can be regarded as the Laplacian matrix of a completely connected undirected graph. Let $0=\lambda_1<\lambda_2\leq\cdots\leq\lambda_m$ be the eigenvalues of $R$ and $0=\mu_1<\mu_2\leq\cdots\leq\mu_m$ be the eigenvalues of $U$. Then, $R-\frac{\lambda_2}{\mu_m}U$ is positive semi-definite.
\end{lem}
Suppose that a multi-agent system has a topology represented by a digraph $G$. The dynamics of each agent is described as
\begin{equation}\label{1}
\begin{array}{ll}
     \dot{x}_i(t)=u_i(t),\ t\geq 0,\ i\in \mathcal{I}_n,
\end{array}
\end{equation}
where $x_i\in R$ is the state of agent $v_i$ and $u_i$ is the control input to be designed. For each agent $v_i,\ i\in\mathcal{I}_n$, assume that the event-triggered time instants are $\{t_l^i,\ l\in N\}$, where $t_0^i=0$ is the initial time and $\{t_l^i,\ l\in N\}\subset\{0,h,2h,\ldots\}$ with $h>0$ being the sampling period. For each agent $v_i,\ i\in\mathcal{I}_n$, define $\hat{x}_i(t)=x_i(t_l^i),\ t_l^i\leq t<t_{l+1}^i$.
For system (\ref{1}) with time delay, the following consensus algorithm is adopted:
\begin{equation}\label{2}
u_i(t)=-\frac{\delta_i}{hd_i}\sum\limits_{j=1}^na_{ij}[\hat{x}_i(t-\tau)-\hat{x}_j(t-\tau)],\ t\geq \tau,
\end{equation}
where $0<\delta_i<\frac{1}{2}$, $d_i$ is the in-degree of agent $v_i$ and $\tau$ is the information communication delay satisfying $0\leq\tau<h$. Note that when $\tau>0$, the control input $u_i(t)$ is chosen as zero for $t\in[0,\tau)$. Another point which should be emphasized is that $u_i(t)$ is also chosen as zero when the agent $v_i$ has no neighbors, i.e., $d_i=0$.
%----------------------------------------------------------------
%----------------------------------------------------------------
\section{Systems without time delay}
We first investigate consensus of (\ref{1}) without time delay, i.e., $\tau=0$. Two kinds of topologies will be considered, respectively.
\subsection{Systems under a strongly connected topology}
In this case, noting that the in-degree matrix $D$ is invertible, the closed-loop system of (\ref{1}) and (\ref{2}) can be summarized as
\begin{equation}\label{3}
\dot{x}(t)=-\frac{1}{h}\Delta D^{-1}L\hat{x}(t),\ t\geq 0,
\end{equation}
where $x=(x_1,\ldots,x_n)^T$, $\hat{x}=(\hat{x}_1,\ldots,\hat{x}_n)^T$, and $\Delta={\rm diag}\{\delta_1,\cdots,\delta_n\}$. Denote $\hat{q}_i=-\sum_{j=1}^nl_{ij}(\hat{x}_i-\hat{x}_j)^2$, $e_i=x_i-\hat{x}_i,\ i\in \mathcal{I}_n$, and $e=(e_1,\ldots,e_n)^T$. The event-triggered time instant $t_l^i$ is defined by the following condition:
\begin{equation}\label{4}
e_i^2(t_l^i)-\sigma_i\frac{1}{4d_i}\hat{q}_i(t_l^i-h)>0,
\end{equation}
where $\sigma_i,\ i\in\mathcal{I}_n$, is a positive constant to be determined later. A direct calculation gets that $\frac{1}{2}\sum_{i=1}^n \xi_i \hat{q}_i=-\frac{1}{2}\sum_{i=1}^n \xi_i\sum_{j=1}^nl_{ij}(\hat{x}_i^2+\hat{x}^2_j-2\hat{x}_i\hat{x}_j)=\sum_{i=1}^n \hat{x}_i\xi_i\sum_{j=1}^nl_{ij}\hat{x}_j$ $=\hat{x}^T\Xi L\hat{x}$, where $\Xi$ is defined as in Lemma \ref{lm.1}.
%----------------------------------------------------------------
\begin{rem}\label{rm.1}
When the condition (\ref{4}) is applied to judge whether an event happens or not at time instant $t_l^i$, the value of $e_i(t)$ obeys the following rule: $e_i(t_l^i)$ is defined as $e_i(t_l^i)=x_i(t_l^i)-\hat{x}_i(t_l^i-h)$ when the judgement is executed and $e_i(t_l^i)=x_i(t_l^i)-\hat{x}_i(t_l^i)$ after the judgement is completed.
\end{rem}
%----------------------------------------------------------------
\begin{thm}\label{th.1}
Suppose that the interaction digraph is strongly connected. System (\ref{1}) without time delay using consensus algorithm (\ref{2}) and event-triggered condition (\ref{4}) will achieve consensus asymptotically for any finite sampling period $h$ if the following condition holds:
\begin{equation}\label{50}
0<\sigma_i<1-2\delta_i,\ i\in\mathcal{I}_n.
\end{equation}
Moreover, $\lim_{t\rightarrow\infty}x(t)=c\mathbf{1}_n$, where $c=\sum_{i=1}^n\delta_i^{-1}d_i\xi_i$ $\times x_i(0)/\sum_{i=1}^n\delta_i^{-1}d_i\xi_i$ and $\xi^T=(\xi_1,\ldots,\xi_n)$ is the positive left eigenvector of $L$ associated with zero eigenvalue satisfying $\sum_{i=1}^n\xi_i$ $=1$ .
\end{thm}
%----------------------------------------------------------------
%----------------------------------------------------------------
\bf Proof. \rm Consider a positive definite quadratic form $V(x)=\frac{1}{2}x^T\Xi D\Delta^{-1}x$, where $\Xi$ is defined as in Lemma \ref{lm.1}. For convenience, denote $V(t)=V(x(t))$, where $x(t)$ is any solution of (\ref{3}). Note that $x(t)=x(lh)-\frac{t-lh}{h}\Delta D^{-1}L\hat{x}(t)$ and $\hat{x}(t)=\hat{x}(lh)$ for $t\in[lh,(l+1)h),\ l\in N$. Deriving $V(t)$ along the trajectories of (\ref{3}) on the interval $[lh,(l+1)h)$, we have
\begin{equation*}
\begin{array}{ll}
\dot{V}(t)=-\frac{1}{h}\hat{x}^T(t)\Xi L\hat{x}(t)+\frac{1}{h}[\hat{x}(t)-x(t)]^T\Xi L\hat{x}(t)\\
\qquad=-\frac{1}{h}\hat{x}^T(t)\Xi L\hat{x}(t)-\frac{1}{h}e^T(lh)\Xi L\hat{x}(t)\\
\qquad\quad+\frac{t-lh}{h^2}\hat{x}^T(t)L^TD^{-1}\Delta\Xi L\hat{x}(t)\\
\qquad=-\frac{1}{2h}\sum\limits_{i=1}^n\xi_i\hat{q}_i(lh)+\frac{1}{h}\sum\limits_{i=1}^n\xi_i\sum\limits_{j=1}^n l_{ij}e_i(lh)[\hat{x}_i(lh)\\
\qquad\quad-\hat{x}_j(lh)]+\frac{t-lh}{h^2}\sum\limits_{i=1}^n\frac{\delta_i\xi_i}{d_i}(\sum\limits_{j=1}^n l_{ij}[\hat{x}_i(lh)-\hat{x}_j(lh)])^2\\
\qquad\leq-\frac{1}{2h}\sum\limits_{i=1}^n\xi_i\hat{q}_i(lh)-\frac{1}{h}\sum\limits_{i=1}^n\xi_i\sum\limits_{j=1,j\neq i}^n l_{ij}\\
\qquad\quad\times\{\frac{1}{4}[\hat{x}_i(lh)-\hat{x}_j(lh)]^2+e_i^2(lh)\}\\
\qquad\quad+\frac{t-lh}{h^2}\sum\limits_{i=1}^n\delta_i\xi_i(\sum\limits_{j=1}^n |l_{ij}|[\hat{x}_i(lh)-\hat{x}_j(lh)]^2)\\
\qquad=-\frac{1}{4h}\sum\limits_{i=1}^n\xi_i\hat{q}_i(lh)+\frac{1}{h}\sum\limits_{i=1}^n\xi_i d_ie_i^2(lh)\\
\qquad\quad+\frac{t-lh}{h^2}\sum\limits_{i=1}^n\delta_i\xi_i \hat{q}_i(lh).
\end{array}
\end{equation*}
Integrating the above inequality from $lh$ to $(l+1)h$, we have $V((l+1)h)\leq V(lh)-\frac{1}{4}\sum_{i=1}^n\xi_i(1-2\delta_i)\hat{q}_i(lh)+\sum\limits_{i=1}^n\xi_id_ie_i^2(lh),\ l\in N.$ Enforcing the event-triggered condition (\ref{4}), we have $V((l+1)h)\leq V(lh)-\frac{1}{4}\sum_{i=1}^n\xi_i(1-2\delta_i)\hat{q}_i(lh)+\frac{1}{4}\sum_{i=1}^n\xi_i\sigma_i\hat{q}_i((l-1)h),\ l\in Z^+.$ Therefore, for $l\in Z^+$, we have
$$
\begin{array}{ll}
V((l+1)h)&\leq V(h)+\frac{1}{4}\sum\limits_{i=1}^n\xi_i\sigma_i\hat{q}_i(0)-\frac{1}{4}\sum\limits_{i=1}^n\xi_i(1-2\delta_i\\
&\quad-\sigma_i)\hat{q}_i(h)-\cdots-\frac{1}{4}\sum\limits_{i=1}^n\xi_i(1-2\delta_i-\sigma_i)\\ &\quad\times\hat{q}_i((l-1)h)-\frac{1}{4}\sum\limits_{i=1}^n\xi_i(1-2\delta_i)\hat{q}_i(lh).
\end{array}
$$
Denote $\gamma_i=1-2\delta_i-\sigma_i$. Note that under condition (\ref{50}), $0< \gamma_i<1-2\delta_i$. In the above inequality, let $l$ approach to infinity, we have
$$
0\leq V(h)+\frac{1}{4}\sum\limits_{i=1}^n\xi_i\sigma_i\hat{q}_i(0) -\frac{1}{4}\sum\limits_{l=1}^\infty\sum\limits_{i=1}^n\gamma_i\xi_i\hat{q}_i(lh).
$$
Noting that $\sum_{i=1}^n\gamma_i\xi_i\hat{q}_i(lh)$ is nonnegative for each $l\in Z^+$, it implies that the series $\sum_{l=1}^\infty\sum_{i=1}^n\gamma_i\xi_i\hat{q}_i(lh)$ is convergent. Hence, $\lim_{l\rightarrow\infty}\sum_{i=1}^n\gamma_i\xi_i\hat{q}_i(lh)=0$, which implies that $\lim_{l\rightarrow\infty}\hat{x}^T(lh)\Xi L\hat{x}(lh)=0$. By Lemma \ref{lm.1}, we have $\lim_{l\rightarrow\infty}\hat{x}^T(lh)U\hat{x}(lh)=0$ and $\hat{x}(lh)\rightarrow\sum_{i=1}^n\xi_i\hat{x}_i(lh)\mathbf{1}_n$ as $l\rightarrow\infty$. Hence, by event-triggered condition (\ref{4}), $\lim_{l\rightarrow\infty}e(lh)=\mathbf{0}_n$. Then, $\lim_{l\rightarrow\infty}Lx(lh)=\lim_{l\rightarrow\infty}Le(lh)+\lim_{l\rightarrow\infty}L\hat{x}(lh)=\mathbf{0}_n$. It implies that $\lim_{t\rightarrow\infty}Lx(t)=\mathbf{0}_n$. Still by Lemma \ref{lm.1}, we have $x(t)\rightarrow a(t)\mathbf{1}_n$ as $t\rightarrow\infty$, where $a(t):[0,\infty)\rightarrow R$. By (\ref{3}), we have $\sum_{i=1}^n\delta_i^{-1}d_i\xi_i\dot{x}_i(t)=0$, which implies that $\sum_{i=1}^n\delta_i^{-1}d_i\xi_ix_i(t)=\sum_{i=1}^n\delta_i^{-1}d_i\xi_ix_i(0)$. This, together with that all the states $x_i(t)$approach to the common value $a(t)$ asymptotically, implies that $a(t)\equiv c$. The proof is complete.
%----------------------------------------------------------------
%----------------------------------------------------------------
\subsection{Systems under a topology with a spanning tree}
In this subsection, we consider the case that the communication digraph contains a spanning tree. Inspired by \cite{2016YiX1}, by reordering the agents, $L$ can be written in the following Perron-Frobenius form:
\begin{equation}\label{71}
L=\left[
  \begin{array}{cccc}
    L^{1,1} & L^{1,2} & \cdots & L^{1,K} \\
    \mathbf{0} & L^{2,2} & \cdots & L^{2,K} \\
    \vdots & \vdots & \ddots & \vdots \\
    \mathbf{0} & \mathbf{0} & \cdots & L^{K,K} \\
  \end{array}
\right],
\end{equation}
where $L^{k,k}\in R^{n_k\times n_k}$ is corresponding to the agents in the $k$th strongly connected component $(SCC)$ of $G$ denoted by $SCC_k,\ k\in\mathcal{I}_K.$ For each $k\in\mathcal{I}_{K-1}$, there exists at least one $j>k$ such that $L^{k,j}\neq\mathbf{0}$.
%----------------------------------------------------------------
\begin{rem}\label{rm.2}
By the following analysis, it is not hard to see that the case of $n_k=1$ for some $k\in\mathcal{I}_{K-1}$ can be regarded as the special case of (\ref{71}). That is, each agent corresponding to $n_k=1$ itself can be regarded as a $SCC$. Moreover, the case of $n_K=1$ can also be regarded as a special case of (\ref{71}). But in this case, since the control input of the corresponding agent is designed as zero, there is no need to consider how to design event-triggered condition and the conclusion in the following theorem still holds.
\end{rem}
%----------------------------------------------------------------
Corresponding to the form of the Laplacian matrix $L$ in (\ref{71}), we use new symbol to denote the state of each agent. That is, let $x_i^k$ be the state of the agent $v_i^k\in SCC_k$, $i\in\mathcal{I}_{n_k}$ and $k\in\mathcal{I}_K$, where $v_i^k$ represents the agent $v_{i+\sum_{j=0}^{k-1}n_j}$ with $n_0=0$. In the following analysis, all the symbols in (\ref{2}) will also be rewritten by default. For each $SCC_k,\ k\in\mathcal{I}_K$, denote $x^k=(x_1^k,\ldots,x_{n_k}^k)^T,\ \hat{x}^k=(\hat{x}_1^k,\ldots,\hat{x}_{n_k}^k)^T$, and $e^k=x^k-\hat{x}^k$, where $\hat{x}_i^k(t)=x_i^k(t^{i+\sum_{j=0}^{k-1}n_j}_l)$ for $t^{i+\sum_{j=0}^{k-1}n_j}_l\leq t<t^{i+\sum_{j=0}^{k-1}n_j}_{l+1}$ and $\{t^{i+\sum_{j=0}^{k-1}n_j}_l,\ l\in N\}$ is the event-triggered time instant sequence. For each $k\in\mathcal{I}_K$, an auxiliary matrix $\tilde{L}^{k,k}$ $=[\tilde{l}_{ij}^{k,k}]_{n_k\times n_k}$ corresponding to $L^{k,k}$ is defined as
$$
\tilde{l}_{ij}^{k,k}=\left\{\begin{array}{ll}
                       l_{ij}^{k,k}, & i\neq j, \\
                       -\sum_{p=1,p\neq i}^{n_k}l^{k,k}_{ip}, & i=j.
                     \end{array}\right.
$$
Let $H^k=L^{k,k}-\tilde{L}^{k,k}={\rm diag}\{h^k_1,\ldots,h^k_{n_k}\},\ k\in\mathcal{I}_K$. We have that each $H^k$, $k\in\mathcal{I}_{K-1}$, is a nonzero positive semi-definite matrix. Specially, $\tilde{L}^{K,K}=L^{K,K}$, i.e., $H^K=\mathbf{0}$. Let $\xi_k^T$ be the left eigenvector of $\tilde{L}^{k,k}$ associated with zero eigenvalue and satisfy $\sum_{i=1}^{n_k}\xi^k_i=1$. Denote $\Xi^k={\mathrm diag}\{\xi_1^k,\ldots,\xi_{n_k}^k\}$ and $\tilde{R}^k=\frac{1}{2}[\Xi^k\tilde{L}^{k,k}+(\tilde{L}^{k,k})^T\Xi^k]$.  Furthermore, denote $\hat{q}_i^k=-\sum_{j=1}^{n_k}\tilde{l}^{k,k}_{ij}(\hat{x}_i^k-\hat{x}_j^k)^2$ for $k\in\mathcal{I}_K$, and $\hat{q}_i^{k,K}=-\sum_{p=k+1}^K\sum_{j=1}^{n_p}l^{k,p}_{ij}(\hat{x}_i^k-\hat{x}_j^p)^2$ for $k\in\mathcal{I}_{K-1}$. Moreover, let $\hat{q}^{k,K}_i=0$ for $k=K$. For the agent $v_i^k\in SCC_k$, $k\in\mathcal{I}_K$, the event-triggered time instant $t_l^{i+\sum_{j=0}^{k-1}n_j}$ is defined by the following condition:
\begin{equation}\label{8}
\begin{array}{ll}
\left[e_i^k\left(t_l^{i+\sum_{j=0}^{k-1}n_j}\right)\right]^2-\sigma_i^k\frac{1}{4d_{i}^k}\left(\hat{q}_i^k\left(t_l^{i+\sum_{j=0}^{k-1}n_j}-h\right)\right.\\
\left.+\hat{q}_i^{k,K}\left(t_l^{i+\sum_{j=0}^{k-1}n_j}-h\right)\right)>0.
\end{array}
\end{equation}
Note that, in essence, (\ref{8}) is merely a restatement of (\ref{4}) for analysis convenience.
%----------------------------------------------------------------

Before going further, we first give a lemma about the relationships among $\hat{q}_i^K$ and $\hat{x}_j^K-\hat{c}$, $i,j\in \mathcal{I}_{n_K}$, where $\hat{c}=\sum_{i=1}^{n_K}\mu_i^K\hat{x}_i^K$ with $\mu_i^K={\delta_i^K}^{-1}d_i^K\xi_i^K/\sum_{i=1}^{n_K}{\delta_i^K}^{-1}d_i^K\xi_i^K$, $i\in \mathcal{I}_{n_K}$.
\begin{lem}\label{lm.2}
For the functions $\hat{q}_i^K$ and $\hat{x}_j^K-\hat{c}$, $i,j\in \mathcal{I}_{n_K}$, given above, the following relationship holds:
$$
(\hat{x}_j^K-\hat{c})^2\leq (n_K-1)\sum\limits_{i=1}^{n_K}\hat{q}_i^K,\ j\in \mathcal{I}_{n_K}.
$$
\end{lem}
\bf Proof. \rm Suppose that not all the states $\hat{x}_i^K$ are equal to $\hat{c}$. Noting that $\hat{c}$ is a weighted average value of all $\hat{x}_i^K$, $i\in \mathcal{I}_{n_K}$, all the agents in $SCC_K$ can be divided into two nonempty groups $A_1$ and $A_2$, where $A_1$ contains the agents whose states are not larger than $\hat{c}$ and $A_2$ contains the rest ones. Since the interaction digraph is strongly connected, for each $v_j^K\in A_1$, there exists $v_k^K\in A_2$ such that there exists a directed path from $v_k^K$ to $v_j^K$ with the form $(v_j^K,v_{i_l}^K),\ldots,(v_{i_1}^K,v_k^K)$. Moreover, we can choose $v_{i_1}^K\in A_1$, i.e., the agent $v_{i_1}^K$ is the closest one in $A_1$ who can receive information from the agent $v_k^K$. It should be emphasized that the inside agents except $v_{i_1}^K$  among this path may belong to $A_2$. Since the largest length of a directed path with no duplicate agents in $SCC_K$ is $n_K-1$, it is easy to get that
$$
\begin{array}{ll}
(\hat{x}_j^K-\hat{c})^2&=(\hat{x}_j^K-\hat{x}_{i_l}^K+\hat{x}_{i_l}^K-\cdots-\hat{x}_{i_1}^K+\hat{x}_{i_1}^K-\hat{c})^2\\
&\leq (|\hat{x}_j^K-\hat{x}_{i_l}^K|+\cdots+|\hat{x}_{i_1}^K-\hat{c}|)^2\\
&\leq (|\hat{x}_j^K-\hat{x}_{i_l}^K|+\cdots+|\hat{x}_{i_1}^K-\hat{x}_k^K|)^2\\
&\leq (n_K-1)[(\hat{x}_j^K-\hat{x}_{i_l}^K)^2+\cdots+(\hat{x}_{i_1}^K-\hat{x}_k^K)^2].
\end{array}
$$
Note that $|\hat{x}_{i_1}^K-\hat{c}|\leq |\hat{x}_{i_1}^K-\hat{x}_k^K|$ plays a key role to get the above inequality. By the form of $\sum_{i=1}^{n_K}\hat{q}_i^K$, we can see that it is the sum of all the squares of the latest event-triggered state errors of the agents who have information communications in $SCC_K$. Hence, the conclusion holds. The case of $v_j^K\in A_2$ can be analyzed similarly.
%----------------------------------------------------------------
%----------------------------------------------------------------

Now, we are ready to establish a theorem which gives sufficient conditions for consensus of (\ref{1}) under a more general topology compared with that in Theorem \ref{th.1}.
\begin{thm}\label{th.2}
Suppose that the interaction digraph contains a spanning tree. System (\ref{1}) without time delay using consensus algorithm (\ref{2}) and event-triggered condition (\ref{8}) will achieve consensus asymptotically for any finite sampling period $h$ if the following condition holds:
\begin{equation}\label{90}
0<\sigma_i^k<1-2\delta_i^k,\ i\in\mathcal{I}_{n_k},\ k\in\mathcal{I}_K.
\end{equation}
Moreover, $\lim_{t\rightarrow\infty}x(t)=c\mathbf{1}_n$, where $c=\sum_{i=1}^{n_K}\mu_i^Kx_i^K(0)$ and $\mu_i^K,\ i\in \mathcal{I}_{n_K},$ is defined as in Lemma \ref{lm.2}.
\end{thm}
%----------------------------------------------------------------
\bf Proof. \rm We only consider the case of $K=2$ since the case of $K>2$ can be tackled similarly. Let $V_1=\frac{1}{2}(x^1-c\mathbf{1}_{n_1})^T\Xi^1 D^1{\Delta^1}^{-1}(x^1-c\mathbf{1}_{n_1})$, where $D^1={\rm diag}\{d_1^1,\ldots,d_{n_1}^1\}$, $\Delta^1={\rm diag}\{\delta_1^1,\ldots,\delta_{n_1}^1\}$. Deriving $V_1$ along the trajectories of the closed-loop system (\ref{1}) and (\ref{2}) on $[lh,(l+1)h)$,  we have
\begin{equation}\label{9}
\begin{array}{ll}
\dot{V}_1(t)&=-\frac{1}{h}[x^1-c\mathbf{1}_{n_1}]^T\Xi^1[L^{1,1}\hat{x}^1+L^{1,2}\hat{x}^2]\\
&=-\frac{1}{h}\{[\hat{x}^1-c \mathbf{1}_{n_1}]+[x^1-\hat{x}^1]\}^T\Xi^1\\
&\quad\times\{L^{1,1}[\hat{x}^1-c\mathbf{1}_{n_1}]+L^{1,2}[\hat{x}^2-c\mathbf{1}_{n_2}]\}\\
&=-\frac{1}{h}[\hat{x}^1-c \mathbf{1}_{n_1}]^T\Xi^1\{L^{1,1}[\hat{x}^1(t)-c\mathbf{1}_{n_1}]\\
&\quad+L^{1,2}[\hat{x}^2-c\mathbf{1}_{n_2}]\}-\frac{1}{h}[e^1(lh)]^T\Xi^1\\
&\quad\times\{L^{1,1}[\hat{x}^1-c\mathbf{1}_{n_1}]+L^{1,2}[\hat{x}^2-c\mathbf{1}_{n_2}]\}\\
&\quad+\frac{t-lh}{h^2}[L^{1,1}\hat{x}^1+L^{1,2}\hat{x}^2]^T{D^1}^{-1}\Delta^1\Xi^1\\
&\quad\times\{L^{1,1} [\hat{x}^1-c\mathbf{1}_{n_1}]+L^{1,2}[\hat{x}^2-c\mathbf{1}_{n_2}]\}.
\end{array}
\end{equation}
Denote the three sum terms of the right side of (\ref{9}) as $F_i(t),\ i=1,2,3$. Next, we analyze them, respectively.
\begin{equation}\label{10}
\begin{array}{rl}
F_1(t)&=-\frac{1}{h}[\hat{x}^1-c \mathbf{1}_{n_1}]^T\Xi^1\tilde{L}^{1,1}[\hat{x}^1-c\mathbf{1}_{n_1}]\\
&\quad-\frac{1}{h}[\hat{x}^1-c \mathbf{1}_{n_1}]^T\Xi^1H^1[\hat{x}^1-c\mathbf{1}_{n_1}]\\
&\quad-\frac{1}{h}[\hat{x}^1-c \mathbf{1}_{n_1}]^T\Xi^1L^{1,2}[\hat{x}^2-c\mathbf{1}_{n_2}]\\
&\leq-\frac{1}{2h}\sum\limits_{i=1}^{n_1}\xi_i^1\hat{q}^1_i(lh)-\frac{1}{h}\sum\limits_{i=1}^{n_1}\xi_i^1h_i^1[\hat{x}_i^1(lh)\\
&\quad-c]^2+\frac{2\kappa}{h} \hat{V}_1(lh)+\frac{1}{h}G_1(lh),
\end{array}
\end{equation}
where $G_1(t)=\frac{1}{4\kappa}\sum_{i=1}^{n_1}\xi_i^1{d_i^1}^{-1}\delta_i^1\{\sum_{j=1}^{n_2}l^{1,2}_{ij}[\hat{x}^2_j-c]\}^2$, $\kappa$ is any positive constant, and $\hat{V}_1(t)=\frac{1}{2}(\hat{x}^1-c\mathbf{1}_{n_1})^T$ $\Xi^1 D^1{\Delta^1}^{-1}(\hat{x}^1-c\mathbf{1}_{n_1})$.
\begin{equation}\label{11}
\begin{array}{ll}
F_2(t)&=-\frac{1}{h}[e^1(lh)]^T\Xi^1\tilde{L}^{1,1}[\hat{x}^1-c\mathbf{1}_{n_1}]\\
&\quad-\frac{1}{h}[e^1(lh)]^T\Xi^1H^1[\hat{x}^1-c\mathbf{1}_{n_1}]\\
&\quad-\frac{1}{h}[e^1(lh)]^T\Xi^1L^{1,2}[\hat{x}^2-c\mathbf{1}_{n_2}]\\
&=\frac{1}{h}\sum\limits_{i=1}^{n_1}\xi^1_i\sum\limits_{j=1}^{n_1} \tilde{l}^{1,1}_{ij}e^1_i(lh)[\hat{x}^1_i(lh)\\
&\quad-\hat{x}^1_j(lh)]-\frac{1}{h}\sum\limits_{i=1}^{n_1}\xi^1_ih_i^1e^1_i(lh)[\hat{x}^1_i(lh)-c]\\
&\quad-\frac{1}{h}\sum\limits_{i=1}^{n_1}\xi^1_i\sum\limits_{j=1}^{n_2}l^{1,2}_{ij}e^1_i(lh)[\hat{x}_j^2(lh)-c]\\
&\leq\frac{1}{4h}\sum\limits_{i=1}^{n_1}\xi^1_i\hat{q}^1_i(lh)+\frac{1}{h}\sum\limits_{i=1}^{n_1}\xi^1_i \tilde{l}^{1,1}_{ii}[e_i^1(lh)]^2\\
&\quad+\frac{1}{h}\sum\limits_{i=1}^{n_1}\xi^1_ih_i^1[e^1_i(lh)]^2+\frac{1}{4h}\sum\limits_{i=1}^{n_1}\xi^1_ih_i^1[\hat{x}^1_i(lh)\\
&\quad-c]^2+\frac{\kappa}{h}\sum\limits_{i=1}^{n_1}\xi^1_id_i^1 [e^1_i(lh)]^2+\frac{1}{h}G_2(lh),
\end{array}
\end{equation}
where $G_2(t)=\frac{1}{4\kappa}\sum_{i=1}^{n_1}\xi_i^1 {d_i^1}^{-1}\{\sum_{j=1}^{n_2}l^{1,2}_{ij}[\hat{x}^2_j-c]\}^2$.
\begin{equation}\label{12}
\begin{array}{ll}
F_3(t)=\frac{t-lh}{h^2}\sum\limits_{i=1}^{n_1}\frac{\xi^1_i\delta_i^1}{d_i^1}\left\{-\sum\limits_{j=1}^{n_1} l^{1,1}_{ij}[\hat{x}^1_i(lh)-\hat{x}^1_j(lh)]\right.\\
\qquad\quad\left.-\sum\limits_{j=1}^{n_2}l^{1,2}_{ij}[\hat{x}_i^1(lh)-\hat{x}_j^2(lh)]\right\}^2\\
\leq\frac{t-lh}{h^2}\sum\limits_{i=1}^{n_1}\xi^1_i\delta_i^1 \left(\hat{q}_i^1(lh)+\hat{q}^{1,2}_i(lh)\right)\leq\frac{t-lh}{h^2}\sum\limits_{i=1}^{n_1}\xi^1_i\delta_i^1\\
\quad\times\left(\hat{q}_i^1(lh)+2h_i^1[\hat{x}^1_i(lh)-c]^2+2G^i(lh)\right),
\end{array}
\end{equation}
where $G^i(lh)=-\sum_{j=1}^{n_2}l^{1,2}_{ij}[\hat{x}_j^2(lh)-c]^2,\ i\in\mathcal{I}_{n_1}$.
By (\ref{9})-(\ref{12}), we have
\begin{equation*}
\begin{array}{ll}
V_1((l+1)h)-V_1(lh)\\
\leq -\frac{1}{4}\sum_{i=1}^{n_1}\xi_i^1\hat{q}^1_i(lh)-\frac{3}{4}\sum\limits_{i=1}^{n_1}\xi^1_ih_i^1[\hat{x}^1_i(lh)-c]^2\\
\quad+\bar{\kappa}\sum\limits_{i=1}^{n_1}\xi^1_id_i^1[e^1_i(lh)]^2+2\kappa \hat{V}_1(lh)+G_1(lh)+G_2(lh)\\
\end{array}
\end{equation*}
\begin{equation}\label{13}
\begin{array}{ll}
\quad+\sum\limits_{i=1}^{n_1}\xi^1_i\delta_i^1\left(\frac{1}{2}\hat{q}_i^1(lh)+h_i^1[\hat{x}^1_i(lh)-c]^2+G^i(lh)\right),
\end{array}
\end{equation}
where $\bar{\kappa}=1+\kappa$. By enforcing the event-triggered condition (\ref{8}), (\ref{13}) implies that, for $l\in Z^+$,
$$
\begin{array}{ll}
V_1((l+1)h)-V_1(lh)\leq-\frac{1}{4}\sum\limits_{i=1}^{n_1}(1-2\delta_i^1)\xi_i^1\hat{q}^1_i(lh)\\
\quad-\frac{1}{4}\sum\limits_{i=1}^{n_1}(3-4\delta_i^1)\xi^1_ih_i^1[\hat{x}^1_i(lh)-c]^2\\
\quad+\frac{\bar{\kappa}}{4}\sum\limits_{i=1}^{n_1}\xi^1_i\sigma_i^1[\hat{q}^1_i((l-1)h)+\hat{q}^{1,2}_i((l-1)h)]\\
\quad+2\kappa \hat{V}_1(lh)+G_1(lh)+G_2(lh)+G_4(lh)\\
\leq-\frac{1}{4}\sum\limits_{i=1}^{n_1}(1-2\delta_i^1)\xi_i^1\left(\hat{q}^1_i(lh)+2h_i^1[\hat{x}^1_i(lh)-c]^2\right)\\
\quad+\frac{\bar{\kappa}}{4}\sum\limits_{i=1}^{n_1}\xi^1_i\sigma_i^1[\hat{q}^1_i((l-1)h)+2h_i^1[\hat{x}^1_i((l-1)h)\\
\quad-c]^2]+2\kappa\hat{V}_1(lh)+G_1(lh)+G_2(lh)\\
\quad+\frac{\bar{\kappa}}{2}G_3((l-1)h)+G_4(lh),
\end{array}
$$
where $G_3(lh)=\sum_{i=1}^{n_1}\xi_i^1\sigma_i^1G^i(lh)$, and $G_4(lh)=\sum_{i=1}^{n_1}\xi_i^1\delta_i^1G^i(lh)$. Therefore, for $l\in Z^+$, we have
\begin{equation}\label{15}
\begin{array}{ll}
V_1((l+1)h)-V_1(h)\\
\leq \frac{\bar{\kappa}}{4}\sum\limits_{i=1}^{n_1}\xi_i^1\sigma_i^1\left(\hat{q}^1_i(0)+2h_i^1[\hat{x}^1_i(0)-c]^2\right)\\
\quad-\frac{1}{4}\sum\limits_{m=1}^{l-1}\sum\limits_{i=1}^{n_1}\xi_i^1(1-2\delta_i^1-\bar{\kappa}\sigma_i^1)[\hat{q}^1_i(mh)+2h_i^1(\hat{x}^1_i(mh)\\
\quad-c)^2]-\frac{1}{4}\sum\limits_{i=1}^{n_1}\xi_i^1(1-2\delta_i^1)[\hat{q}^1_i(lh)+2h_i^1(\hat{x}^1_i(lh)\\
\quad-c)^2]+\frac{\bar{\kappa}}{2}\sum\limits_{m=0}^{l-1}G_3(mh)+\sum\limits_{m=1}^{l}[2\kappa\hat{V}_1(mh)\\
\quad+G_1(mh)+G_2(mh)+G_4(mh)].
\end{array}
\end{equation}
Together with $\frac{1}{2}\sum_{i=1}^{n_1} \xi_i^1 \hat{q}^1_i(lh)=[\hat{x}^1(lh)]^T\Xi^1 \tilde{L}^{1,1}\hat{x}^1(lh)$ $=[\hat{x}^1(lh)-c\mathbf{1}_{n_1}]^T\tilde{R}^1[\hat{x}^1(lh)-c\mathbf{1}_{n_1}]$, it is easy to get that $\sum_{i=1}^{n_1} \xi_i^1 (\frac{1}{2}\hat{q}^1_i(lh)+h_i^1[\hat{x}^1_i(lh)-c]^2)=[\hat{x}^1(lh)-c\mathbf{1}_{n_1}]^TR^1[\hat{x}^1(lh)-c\mathbf{1}_{n_1}]$, where $R^1=(1/2)[\Xi^1L^{1,1}+(L^{1,1})^T\Xi^1]$. Define $\alpha=\frac{\xi^1_{\max}d^1_{\max}}{2\delta_{min}^1\lambda_1(R^1)}$ with $\delta_{min}^1=\min_{1\leq i\leq n_1}\{\delta_i^1\}$, $\xi^1_{\max}=\max_{1\leq i\leq n_1}\{\xi_i^1\}$ and $d^1_{\max}=\max_{1\leq i\leq n_1}\{d_i^1\}$.  Under condition (\ref{90}), we can choose $\kappa$ to satisfy $\gamma_i^1\triangleq1-2\delta_i^1-\bar{\kappa}\sigma_i^1>4\kappa\alpha$. From (\ref{15}), let $l$ approach to infinity, we have
\begin{equation}\label{151}
\begin{array}{ll}
0\leq V_1(h)+\frac{\bar{\kappa}}{4}\sum\limits_{i=1}^{n_1}\xi_i^1\sigma_i^1\left(\hat{q}^1_i(0)+2h_i^1[\hat{x}^1_i(0)-c]^2\right)\\
\qquad-\frac{1}{2}(\gamma_{min}^1-4\kappa\alpha)\sum\limits_{m=1}^{\infty}[\hat{x}^1(mh)-c\mathbf{1}_{n_1}]^T R^1\\
\qquad\times[\hat{x}^1(mh)-c\mathbf{1}_{n_1}]+\frac{1}{2}\bar{\kappa}\sum\limits_{m=0}^{\infty}G_3(mh)\\
\qquad+\sum\limits_{m=1}^{\infty}[G_1(mh)+G_2(mh)+G_4(mh)],
\end{array}
\end{equation}
where $\gamma_{min}^1=\min_{1\leq i\leq n_1}\{\gamma_i^1\}$. Note that $\hat{V}_1(t)\leq\alpha (\hat{x}^1-c\mathbf{1}_{n_1})^TR^1(\hat{x}^1-c\mathbf{1}_{n_1})$ has been used to get the above inequality. Define $a(t)=\sum_{i=1}^{n_2}\mu_i^2x_i^2(t)$. It is easy to get that $a(t)\equiv c$. By the event-triggered condition (\ref{8}), we have
\begin{equation*}
\begin{array}{ll}
[\hat{c}(mh)-c]^2&=[\hat{c}(mh)-a(t)]^2\\
&\leq n_2\sum\limits_{i=1}^{n_2}(\mu_i^2)^2[x_i^2(mh)-\hat{x}_i^2(mh)]^2\\
&\leq n_2\sum\limits_{i=1}^{n_2}(\mu_i^2)^2\frac{\sigma_i^2}{4d_i^2}\hat{q}_i^2(mh-h),
\end{array}
\end{equation*}
where $\hat{c}(t)$ is defined as in Lemma \ref{lm.2} and $m\in Z^+$. By Lemma \ref{lm.2}, it follows that, for each $\ j\in\mathcal{I}_{n_2}$,
\begin{equation*}
\begin{array}{ll}
[\hat{x}_j^2(mh)-c]^2&\leq 2[(\hat{x}_j^2(mh)-\hat{c}(mh))^2+(\hat{c}(mh)-c)^2]\\
&\leq 2[(n_2-1)\sum\limits_{i=1}^{n_2}\hat{q}_i^2(mh)\\
&\quad+n_2\sum\limits_{i=1}^{n_2}(\mu_i^2)^2\frac{\sigma_i^2}{4d_i^2}\hat{q}_i^2(mh-h)],
\end{array}
\end{equation*}
where $m\in Z^+$. From the proof of Theorem \ref{th.1}, we know that the series $\sum_{m=1}^\infty\sum_{i=1}^{n_2}\hat{q}_i^2(mh)$ is convergent. This, together with the above inequality, implies that all the series $\sum_{m=1}^{\infty}G_i(mh),\ 1\leq i\leq 4,$ are convergent. Then, by (\ref{151}), we have that the series $\sum_{m=1}^{\infty}\left[\hat{x}^1(mh)-c\mathbf{1}_{n_1}]^TR^1[\hat{x}^1(mh)-c\mathbf{1}_{n_1}\right]$ is also convergent. Since $R^1$ is positive definite, it follows that $\lim_{m\rightarrow\infty}\hat{x}^1(mh)=c\mathbf{1}_{n_1}$. By the event-triggered condition (\ref{8}), $\lim_{m\rightarrow\infty}e^1(mh)=\mathbf{0}_{n_1}$. Hence, $\lim_{m\rightarrow\infty}x^1(mh)=\lim_{m\rightarrow\infty}e^1(mh)+\hat{x}^1(mh)=c\mathbf{1}_{n_1}$. Also, from the proof of Theorem \ref{th.1}, we have $\lim_{m\rightarrow\infty}\hat{x}^2(mh)=c\mathbf{1}_{n_2}$. Noting that $x^1(t)=x^1(mh)-\frac{t-mh}{h}\Delta^1{D^1}^{-1}[L^{1,1}\hat{x}^1(mh)+L^{1,2}\hat{x}^2(mh)]$ for $t\in[mh,(m+1)h)$, it follows that $\lim_{t\rightarrow\infty}x^1(t)=c\mathbf{1}_{n_1}$. Moreover, it follows from Theorem 1 that $\lim_{t\rightarrow\infty}x^2(t)=c\mathbf{1}_{n_2}$. Hence, the conclusion holds and the proof is complete.
%======================================================================
\section{Systems with time delay}
%----------------------------------------------------------------
In this section, we investigate consensus of (\ref{1}) with time delay. Two kinds of topologies will also be considered, respectively.
\subsection{Systems under a strongly connected topology}
The closed-loop system (\ref{1}) and (\ref{2}) can be summarized as
\begin{equation}\label{19}
\dot{x}(t)=
\left\{
\begin{array}{ll}
0,\ t\in[0,\tau),\\
-\frac{1}{h}\Delta D^{-1}L\hat{x}(t-\tau),\ t\geq\tau.
\end{array}
\right.
\end{equation}

Still by using event-triggered condition (\ref{4}), we have the following theorem.
%----------------------------------------------------------------
\begin{thm}\label{th.3}
Suppose that the interaction digraph is strongly connected. System (\ref{1}) with time delay using consensus algorithm (\ref{2}) and event-triggered condition (\ref{4}) will achieve consensus asymptotically for any finite sampling period $h$ if the following conditions hold:
\begin{equation}\label{20}
0<\sigma_i<1-2\delta_i\ \mbox{and}\ 0<\tau<h\beta,
\end{equation}
where $\beta=\min\{1,\frac{1-2\delta_i-\sigma_i}{4\delta_i}, i\in\mathcal{I}_n\}$. Moreover, $\lim\limits_{t\rightarrow\infty}x(t)$ $=c\mathbf{1}_n$, where $c=\sum_{i=1}^n\delta_i^{-1}d_i\xi_ix_i(0)/\sum_{i=1}^n\delta_i^{-1}d_i\xi_i$ and $\xi^T=(\xi_1,\ldots,\xi_n)$ is the positive left eigenvector of $L$ associated with zero eigenvalue satisfying $\sum_{i=1}^n\xi_i$ $=1$.
\end{thm}
%----------------------------------------------------------------
%----------------------------------------------------------------
\bf Proof. \rm Consider the same positive definite quadratic form as in the proof of Theorem \ref{th.1}. Note that $x(t)=x(lh)-\frac{\tau}{h}\Delta D^{-1}L\hat{x}((l-1)h)-\frac{t-lh-\tau}{h}\Delta D^{-1}L\hat{x}(lh)$ for $t\in[lh+\tau,(l+1)h+\tau],\ l\in N$. Deriving $V(t)$ along the trajectories of (\ref{19}) on the interval $[lh+\tau,(l+1)h+\tau)$, we have
\begin{equation*}
\begin{array}{ll}
\dot{V}(t)=-\frac{1}{h}\hat{x}^T(lh)\Xi L\hat{x}(lh)+\frac{1}{h}[\hat{x}(lh)-x(t)]^T\Xi L\hat{x}(lh)\\
=-\frac{1}{h}\hat{x}^T(lh)\Xi L\hat{x}(lh)-\frac{1}{h}e^T(lh)\Xi L\hat{x}(lh)\\
\quad+\frac{\tau}{h^2}\hat{x}^T((l-1)h)L^TD^{-1}\Delta\Xi L\hat{x}(lh)\\
\quad+\frac{t-lh-\tau}{h^2}\hat{x}^T(lh)L^TD^{-1}\Delta\Xi L\hat{x}(lh)\\
\leq-\frac{1}{2h}\sum\limits_{i=1}^n\xi_i\hat{q}_i(lh)+\frac{1}{h}\sum\limits_{i=1}^n\xi_i\sum\limits_{j=1}^n l_{ij}e_i(lh)[\hat{x}_i(lh)-\hat{x}_j(lh)]\\
\quad+\frac{\tau}{2h^2}\sum\limits_{i=1}^n\frac{\delta_i\xi_i}{d_i}(\sum\limits_{j=1}^n l_{ij}[\hat{x}_i((l-1)h)-\hat{x}_j((l-1)h)])^2\\
\quad+\frac{t-lh-\tau/2}{h^2}\sum\limits_{i=1}^n\frac{\delta_i\xi_i}{d_i}(\sum\limits_{j=1}^n l_{ij}[\hat{x}_i(lh)-\hat{x}_j(lh)])^2\\
\leq-\frac{1}{2h}\sum\limits_{i=1}^n\xi_i\hat{q}_i(lh)-\frac{1}{h}\sum\limits_{i=1}^n\xi_i\sum\limits_{j=1,j\neq i}^n l_{ij}\\
\quad\times\{\frac{1}{4}[\hat{x}_i(lh)-\hat{x}_j(lh)]^2+e_i^2(lh)\}\\
%\end{array}
%\end{equation*}
%\begin{equation*}
%\begin{array}{ll}
\quad+\frac{\tau}{2h^2}\sum\limits_{i=1}^n\delta_i\xi_i(\sum\limits_{j=1}^n |l_{ij}|[\hat{x}_i((l-1)h)-\hat{x}_j((l-1)h)]^2)\\
\quad+\frac{t-lh-\tau/2}{h^2}\sum\limits_{i=1}^n\delta_i\xi_i(\sum\limits_{j=1}^n |l_{ij}|[\hat{x}_i(lh)-\hat{x}_j(lh)]^2)\\
=-\frac{1}{4h}\sum\limits_{i=1}^n\xi_i\hat{q}_i(lh)+\frac{1}{h}\sum\limits_{i=1}^n\xi_i d_ie_i^2(lh)\\
\quad+\frac{\tau}{2h^2}\sum\limits_{i=1}^n\delta_i\xi_i \hat{q}_i((l-1)h)+\frac{t-lh-\tau/2}{h^2}\sum\limits_{i=1}^n\delta_i\xi_i \hat{q}_i(lh).
\end{array}
\end{equation*}
Integrating the above inequality from $lh+\tau$ to $(l+1)h+\tau$, we have $V((l+1)h+\tau)\leq V(lh+\tau)-\frac{1}{4}\sum_{i=1}^n\xi_i[1-2(1+\frac{\tau}{h})\delta_i]\hat{q}_i(lh)+\frac{\tau}{2h}\sum_{i=1}^n\delta_i\xi_i \hat{q}_i((l-1)h)+\sum_{i=1}^n\xi_i d_ie_i^2(lh),\ l\in Z^+.$ Enforcing event-triggered condition (\ref{4}), we have $V((l+1)h+\tau)\leq V(lh+\tau)-\frac{1}{4}\sum\limits_{i=1}^n\xi_i[1-2(1+\frac{\tau}{h})\delta_i]\hat{q}_i(lh)+\frac{1}{4}\sum\limits_{i=1}^n\xi_i(\sigma_i+\frac{2\tau}{h}\delta_i)\hat{q}_i((l-1)h),\ l\in Z^+.$ Therefore, for $l\in Z^+$, we have
$$
\begin{array}{ll}
V((l+1)h+\tau)\leq V(h+\tau)+\frac{1}{4}\sum\limits_{i=1}^n\xi_i(\sigma_i+\frac{2\tau}{h}\delta_i)\hat{q}_i(0)\\ \quad-\frac{1}{4}\sum\limits_{i=1}^n\xi_i[1-2(1+\frac{2\tau}{h})\delta_i-\sigma_i]\hat{q}_i(h)-\cdots\\
\quad-\frac{1}{4}\sum\limits_{i=1}^n\xi_i[1-2(1+\frac{2\tau}{h})\delta_i-\sigma_i]\hat{q}_i((l-1)h)\\ \quad-\frac{1}{4}\sum\limits_{i=1}^n\xi_i[1-2(1+\frac{\tau}{h})\delta_i]\hat{q}_i(lh).
\end{array}
$$
The rest is just as same as that in the proof of Theorem \ref{th.1} and is omitted for space saving.
%----------------------------------------------------------------
\subsection{Systems under a topology with a spanning tree}
%----------------------------------------------------------------
When system (\ref{1}) has a communication digraph which contains a spanning tree, a similar analysis as that in Subsection 3.2 can be given. Actually, by comparing the proof of Theorem \ref{th.3} with that of Theorem \ref{th.1}, the derivative of the same $V_1$ as that in the proof of Theorem \ref{th.2} along the trajectories of the closed-loop system (\ref{1}) and (\ref{2}) can be obtained by changing the coefficient of $F_3(t)$ in (\ref{9}) to be $\frac{t-lh-\tau}{h^2}$ and adding a function $F_4(t)=\frac{\tau}{h^2}[L^{1,1}\hat{x}^1((l-1)h)+L^{1,2}\hat{x}^2((l-1)h){D^1}^{-1}\Delta^1\Xi^1\{L^{1,1} [\hat{x}^1(t)-\hat{c}\mathbf{1}_{n_1}]+L^{1,2}[\hat{x}^2(t)-\hat{c}\mathbf{1}_{n_2}]\}$. Then, the following theorem about system (\ref{1}) with time delay under a topology with a spanning tree can be obtained. Since the proof is not too complicated, we omit it for space saving.
%----------------------------------------------------------------
\begin{thm}\label{th.4}
Suppose that the interaction digraph contains a spanning tree. System (\ref{1}) with time delay using consensus algorithm (\ref{2}) and event-triggered condition (\ref{8}) will achieve consensus asymptotically for any finite sampling period $h$ if the following conditions hold:
\begin{equation}\label{21}
0<\sigma_i^k<1-2\delta_i^k\ \mbox{and}\ 0<\tau<h\beta,
\end{equation}
where $\beta=\min\left\{1,\frac{1-2\delta_i^k-\sigma_i^k}{4\delta_i^k}, i\in\mathcal{I}_{n_k},\ k\in\mathcal{I}_K\right\}.$ Moreover, $\lim\limits_{t\rightarrow\infty}x(t)=c\mathbf{1}_n$, where $c=\sum_{i=1}^{n_K}{\delta_i^K}^{-1}d_i^K\xi_i^Kx_i^K(0)$ $/\sum_{i=1}^{n_K}{\delta_i^K}^{-1}d_i^K\xi_i^K$ and $\xi_K^T=(\xi^K_1,\ldots,\xi^K_{n_K})$ with $\sum_{i=1}^{n_K}\xi_i^K=1$ is the positive left eigenvector of $L^{K,K}$ associated with zero eigenvalue.
\end{thm}
%----------------------------------------------------------------
\begin{rem}\label{rm.3}
A common merit of Theorems 1-4 is that the sampling period can be arbitrarily large. This point is obviously superior to the existing results which demand that the sampling period should be bounded by certain number dependent on the spectrum of Laplacian matrix, which is actually global information. One might think that, as a distributed control design, it is a drawback that all the agents need to share a common sampling period. Just as all the agents share a common form of consensus algorithm, to share a common sampling period  can be regarded as a prerequisite. The true distributed control should be reflected by the neighbor-based information sharing during the state evolution of the whole system.
\end{rem}
%----------------------------------------------------------------
\begin{rem}\label{rm.4}
In Theorem \ref{th.3}, the time delay $\tau$ has a bound dependent on both the sampling period $h$ and the parameters $\delta_i$ and $\sigma_i$ of each agent, which can be regarded as the using of global information. Noting that $\lim_{\delta_i\rightarrow 0}\frac{1-2\delta_i-\sigma_i}{4\delta_i}=\infty$, $\tau$ actually can be bounded only by $h$ if $\delta_i$ and $\sigma_i$ are chosen in advance to satisfy $\frac{1-2\delta_i-\sigma_i}{4\delta_i}\geq 1$. A similar conclusion holds for the delay in Theorem \ref{th.4}. This point may result in a very useful application. That is, if the range of the time delay is known, a large enough sampling period $h$ can be chosen in (\ref{2}) to guarantee the achievement of consensus for system (\ref{1}).
\end{rem}
%======================================================================
\section{Simulation example}
\begin{figure}[!htb]
\centering
\scalebox{0.6}[0.6]{\includegraphics{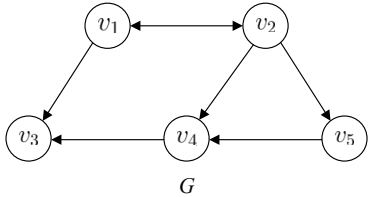}}
\caption{Digraph $G$ which represents the topology}
\label{Fig.1}
\end{figure}

In this section, we give a numerical example to illustrate the effectiveness of the obtained theoretical results. The considered system consists of five agents. Figure \ref{Fig.1} shows the interaction digraph $G$ which represents the system's topology. Obviously, $G$ contains a spanning tree. For simplicity, we only consider the case with time delay.
\begin{exmp}\label{ex:1}
Let the initial states of system (\ref{1}) be $x(0)=[19,5,1,-8,-4]^T.$ We choose the parameters $\delta_1=\delta_3=\delta_5=1/4$, $\delta_2=\delta_4=1/5$, the event-checking period $h=0.1$. Moreover, for simplicity, a common value 1/5 for all the parameters $\sigma_i,\ i\in\mathcal{I}_n$, and $\tau=0.02$ are chosen to satisfy the condition (\ref{21}). By Theorem \ref{th.4}, system (\ref{1}) using (\ref{2}) driven by the event-triggered condition (\ref{8}) can achieve consensus asymptotically and the final consensus state is $101/9\approx 11.22$. The time instants when the events occur for each agent are shown in Figure \ref{Fig.2}. Figure \ref{Fig.3} shows the states and the convergence tendencies of all the agents, which are consistent with the conclusion of Theorem \ref{th.4}.
\begin{figure}[!htb]
\centering
\scalebox{0.6}[0.6]{\includegraphics{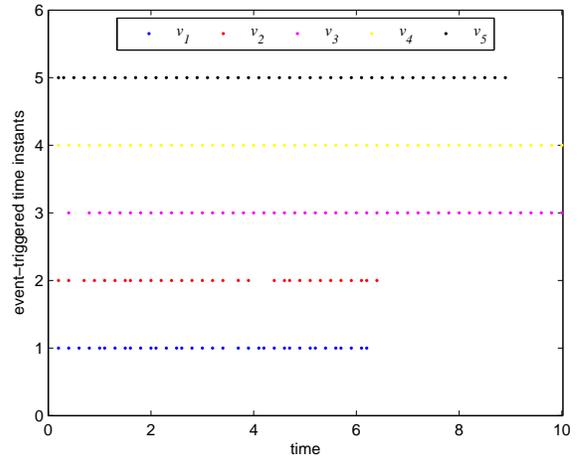}}
\caption{Event-triggered time instants under (\ref{8})}
\label{Fig.2}
\end{figure}
\begin{figure}[!htb]
\centering
\scalebox{0.6}[0.6]{\includegraphics{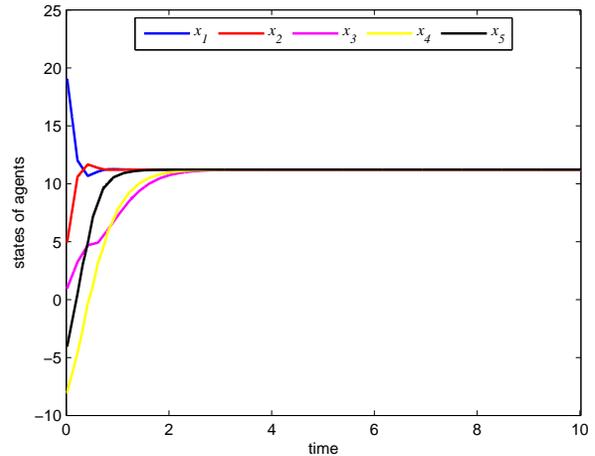}}
\caption{States of (\ref{1}) using (\ref{2}) under topology $G$}
\label{Fig.3}
\end{figure}
\end{exmp}
%======================================================================
\section{Conclusion}
In this paper, we have investigated the consensus problem of first-order multi-agent systems under directed topology which contains a spanning tree. Only the data at periodic time instants are used for event judgements. Both the case without delay and the case with delay have been investigated. Under the designed consensus algorithm, the sampling period can be arbitrarily large. For the case with delay, the delay is only needed to be bounded by the sampling period to guarantee consensus. To investigate the consensus problem of multi-agent systems with time-varying delays and/or under switching topologies will be our future work.
%======================================================================
%\vspace{-0.1cm}
\section*{Acknowledgement}
This work was supported by NSFC (Nos. 61873136, 61374062 and 61075114), Science Foundation of Shandong Province for Distinguished Young Scholars (No. JQ201419), Natural Science Foundation of Shandong Province (No. ZR2015FM023), Postdoctoral Science Foundation of China (No. 2015M571995) and Postdoctoral Application Research Project of Qingdao.
%======================================================================
%%\bibliographystyle{D:/BIB/bibliographystyle/IEEEtran}
%%\bibliographystyle{D:/BIB/bibliographystyle/IEEEtranS}
%\bibliographystyle{D:/BIB/bibliographystyle/elsart-num}
%\bibliographystyle{D:/BIB/bibliographystyle/elsart-harv}
%%\bibliographystyle{D:/BIB/bibliographystyle/elsart-num-names}
%%\bibliographystyle{D:/BIB/bibliographystyle/elsart-num-sort}
%%\bibliographystyle{D:/BIB/bibliographystyle/latex8}
\bibliographystyle{plain}        % Include this if you use bibtex
\bibliography{reference}

\end{document}